\documentclass[10pt]{blois07}
\usepackage{graphicx,epsfig}
\usepackage{cite,./mcite}
\newcommand{\ccc}[1]{} 

\begin{document}
\title{Blois07/EDS07 \\ Proceedings}
\author{{{Vojt\v{e}ch Kundr\'{a}t}}, Jan Ka\v{s}par, 
Milo\v{s} Lokaj\'{i}\v{c}ek}
\institute{Institute of Physics of the AS CR, v. v. i., 182 21 Prague 8, 
Czech Republic}
\maketitle{\bf{To the theory of high-energy elastic nucleon collisions}}
\begin{abstract}
The commonly used West and Yennie integral formula for 
the relative phase between the Coulomb and elastic hadronic 
amplitudes requires for the phase of the elastic hadronic 
amplitude to be constant at all kinematically allowed 
values of $t$. More general interference formula based on 
the eikonal model approach does not exhibit such limitation.
The corresponding differences will be demonstrated and some
predictions of different phenomenological models for elastic
pp scattering at energy of 14 TeV at the LHC will be given. 
Special attention will be devoted to determination of luminosity 
from elastic scattering data; it will be shown that the
systematic error might reach till 5 {\%} if the luminosity 
is derived from the values in the center of the interference 
region with the help of West and Yennie formula.  
\end{abstract}
\section{Limited validity of West and Yennie integral formula}
\label{sec1}
It has been shown in our earlier papers (see \cite{kun1}  
and \cite{vrko}) that the integral formula of West and Yennie 
\cite{west} for the real relative phase between Coulomb and
hadronic amplitudes 
\begin{equation}
\alpha \Phi (s,t) = \mp \alpha \bigg [ \ln \bigg ( {{-t} \over s} \bigg ) -
\int_{-4 p^2}^{0} {{ d \tau} \over {|t - \tau|}} \bigg ( 
1 - {{F^N(s,\tau)} \over {F^N(s,t)}} \bigg ) \bigg ]
\label{wy1}
\end{equation}
requires for the hadronic amplitude $F^N(s,t)$ to have 
the constant phase at any kinematically allowed value
of $t$; $s$ being the value of the total CMS energy,
$t$ the four momentum transfer squared, $p$ the value
of the CMS momentum and $\alpha = 1/137.036$ the fine 
structure constant. The upper (lower) sign corresponds 
to the pp ($\mathrm{\bar{p}p}$) scattering.
It follows in such a case
\begin{equation}
\int_{-4 p^2}^{0} {{ d \tau} \over {|t - \tau|}} \Im \bigg ( 
{{F^N(s,\tau)} \over {F^N(s,t)}} \bigg ) \; \equiv \; 0
\label{wy2}
\end{equation}
and further 
\begin{equation}
 \int\limits_{-4 p^2}^{0} {{d \tau} \over {|t - \tau|}}
\bigl [ \Re F^N(s,t) \Im F^N(s,\tau) - \Re F^N(s,\tau) \Im F^N(s,t) \bigr ]
\; \equiv \; 0.
\label{wy3}
\end{equation}
Introducing then 
\begin{equation}
F^{N}(s,t) \; = \; i |F^N(s,t)| e^ {-i \zeta^N (s,t)},
\label{wy4}
\end{equation}
it is possible to write further
\begin{equation}
\int\limits_{-4p^2}^{t} d \tau \;
{{\sin[\zeta^N(s,t) - \zeta^N(s,\tau)]} \over{t - \tau}} \;  |F^N(s,\tau)|
- \int\limits_{t}^{0} d \tau \;
{{\sin[\zeta^N(s,t) - \zeta^N(s,\tau)]}\over{t - \tau}} \;  |F^N(s,\tau)|
\; \equiv \; 0
\label{nu1}
\end{equation}
for any $t \in [-4p^2, 0]$.
Both the integrals in Eq.~(\ref{nu1}) are
proper integrals provided the first derivative
of the hadronic phase $[\zeta^N(s,t)]'$ according to $t$
variable is finite. It is evident that Eq.~(\ref{nu1})
is fulfilled if the phase is $t$ independent, i.e., if
\begin{equation}
\zeta^N(s,t) \; = \; \zeta^N(s,\tau) \; \equiv \; \zeta^N(s). 
\label{nu2}
\end{equation}
It has been shown in Ref.~\cite{vrko} that Eq.~(\ref{nu2})
represents the unique solution of Eq.~(\ref{nu1}), if the 
relative phase between the Coul. and hadr. amplitudes 
is to be a real quantity as commonly assumed. 

The problem of the $t$ independence of hadronic phase 
was mentioned for the first time in Ref. \cite{amal}.
However, this independence was used as an assumption 
only. Adding the other important
assumption concerning purely exponential $t$ 
dependence of the modulus of hadronic amplitude 
{\it{in the whole kinematically allowed region of $t$}}
it was possible to perform analytically the integration in 
Eq.~(\ref{wy1}) (accepting also other approximations 
valid at asymptotic energies only - for detail see 
Ref. \cite{kun1}). For the total elastic scattering amplitude 
the simplified West and Yennie formula
\begin{equation}
F^{C+N}(s,t) =
\pm {\alpha s \over t} f_1(t)f_2(t)e^{i\alpha \Phi}+
{\sigma_{tot} \over {4\pi}} p\sqrt {s} (\rho+i)e^{Bt/2}
\label{wy5}
\end{equation}
was then obtained.
Here $f_1(t)$ and $f_2(t)$ are the dipole form factors
(added by hand only), $B$ is the constant diffractive slope, 
$\sigma_{tot}$ the value of the total cross section and 
the constant $\rho = {\Re F^N(s,0)} / {\Im F^N(s,0)}$. 
These three quantities may depend on $s$ only. The
relative phase  $\alpha \Phi(s,t)$ exhibits then  
logarithmic $t$ dependence
\begin{equation}
\alpha \Phi (s,t) = \mp \alpha \bigg [ \ln \bigg ({{-B t} \over 2} 
\bigg ) + \gamma \bigg ],
\label{wy6}
\end{equation}
where $\gamma = 0.577215$ is Euler's constant.
\section{General eikonal model approach}
\label{sec2}
The contemporary experimental data as well
as the phenomenological models of high-energy 
elastic nucleon scattering show, however, convincingly
that the quantity $\rho$ cannot be $t$ independent. 
Consequently, the West and Yennie approach \cite{west} 
is not a convenient tool for description of interference 
between the Coulomb and elastic hadronic interactions
of charged nucleons. However, the approach based 
on the eikonal model removes such troubles.
The general formula for the total elastic scattering 
amplitude proposed in Ref. \cite{kun2} may be
valid at any $s$ and $t$; it may be written as 
\begin{equation}
F^{C+N}(s,t) = \pm {\alpha s\over t}f_1(t)f_2(t) +
F^{N}(s,t)\Bigg [1\mp i\alpha G(s,t) \Bigg ],  
\label{kl1}
\end{equation}
where
\begin{equation}
\!\!\!\!\!\!\!G(s,t) = \int\limits_{-4p^2}^0
dt'\Bigg \{ \ln \bigg( {t'\over t} \bigg )
{d \over{dt'}}
\bigg[f_1(t')f_2(t')\bigg]
+ {1\over {2\pi}}\Bigg [{F^{N}(s,t')\over F^{N}(s,t)}-1\Bigg]
I(t,t')\Bigg \},
\label{kl2}
\end{equation}
and
\begin{equation}
I(t,t')=\int\limits_0^{2\pi}d{\Phi^{\prime \prime}}
{f_1(t^{\prime \prime})f_2(t^{\prime \prime})\over t^{\prime \prime}}, \;\;
t^{\prime \prime}=t+t'+2\sqrt{tt'}\cos{\Phi}^{\prime \prime}.
\label{kl3}
\end{equation}
Instead of the $t$ independent quantities
$B$ and $\rho$, it is now necessary to consider
$t$ dependent quantities being defined as
\begin{equation}
B(s,t)= {d\over {dt}}\bigg[\ln {d \sigma^{N}\over {dt}}\bigg] =
{2\over |F^{N}(s,t)|}{d\over {dt}}|F^{N}(s,t)|, \;\;\;\;\;\;\;\;
\rho (s,t) = {{\Re F^{N}(s,t)} \over {\Im F^{N}(s,t)}}.
\label{sl1}
\end{equation}
The total cross section is then given with the help 
of the optical theorem as
\begin{equation}
\sigma_{tot} (s) = {{4 \pi}\over {p \sqrt{s}}} \Im F^{N}(s,t=0).
\label{ot1}
\end{equation}
\section{Experimental data and West and Yennie formula}
\label{sec3}
It is the differential cross section that is
determined in corresponding experiments. In
our normalization it equals 
\begin{equation}
{{d \sigma (s,t)} \over {dt}}= {\pi\over {sp^2}}|F^{C+N}(s,t)|^2.
\label{ds1}
\end{equation}
In the past practically in all actual experiments the 
simplified West and Yennie elastic amplitude defined 
by Eqs.~(\ref{wy5}) - (\ref{wy6}) has been used for the
analysis of data at $|t| \le 0.01$ GeV$^2$, in spite 
of the fact that the theoretical assumptions under
which the amplitude was derived are not fulfilled
at all kinematically allowed values of $t$ but only
in its narrow region in forward direction. Generally,
some important discrepancies exist.
The analysis of corresponding behavior has been
performed in Ref. \cite{proc} where the general formula 
(Eqs.~(\ref{kl1}) - ({\ref{kl3})) has been applied to
data for pp scattering at energy of 53 GeV under different
limitations of some free parameters specifying the
modulus and the phase of the hadronic amplitude 
(as used in Ref. \cite{kun2} - Eqs.~(40) and (42)). 
The corresponding results were derived in
Ref. \cite{proc}) and are represented in Fig. 1. 
First, only the phase has been fixed by putting
$\tan \zeta^N(t) = \rho = 0.077$ according to the 
earlier fit \cite{amos} (based on West and Yennie
formula), while the modulus parameters have
been fitted. This fit has been compared to the 
case when only $\rho$ has been assumed to be constant,
but free; the optimum in such a case has been obtained 
with $\rho = 0.065$. In both the cases the experimental 
data have been represented by the square of the modulus
fitted in the whole measured interval; see the solid 
line in Fig. 1. In addition to, in the
other case the modulus formula has been limited
to a simple exponential form (with two free parameters). 
Fundamental differences from experimental data
exist now; only unsubstantial difference being
obtained when $\rho$ has been fixed ($\rho = 0.077$)
and when it has been left free and fitted to 
$\rho = 0.021$; see dotted and dashed lines in Fig. 1.
It is evident that the assumption of constant 
diffractive slope $B$ is in strong contradiction 
to the experimental data. The curves corresponding
to the fits with
constant and $t$ dependent quantities 
$\rho$ are nearly the same, if the modulus is fitted 
and only weak dependence of $\rho$ on $t$ is allowed. 
Better $\chi^2$ quantity is obtained if strong dependence 
on $t$ is allowed.
\begin{center}
\begin{table}[t]
\begin{tabular}{|c|c|c|c|c|c|c|c|}  
\cline{1-8} 
    & & & & & & & \\[-0.3cm]
   model & $\sigma_{tot}$ & $\sigma_{el}$ & $B(0)$ & $\rho$
   & $\sqrt{<b^2_{tot}>}$ & $\sqrt{<b^2_{el}>}$ & $\sqrt{<b^2_{inel}>}$ \\ 
   & [mb] & [mb] & [GeV$^{-2}$] &  & [fm] & [fm] & [fm] \\
\cline{1-8} 
   Islam          & 109.17 & 21.99 & 31.43 & 0.123 & 1.552 & 1.048 & 1.659 \\
Petrov et al.2P&  94.97 & 23.94 & 19.34 & 0.097 & 1.227 & 0.875 & 1.324 \\
Petrov et al.3P& 108.22 & 29.70 & 20.53 & 0.111 & 1.263 & 0.901 & 1.375 \\
Bourrely et al.& 103.64 & 28.51 & 20.19 & 0.121 & 1.249 & 0.876 & 1.399 \\
Block et al.   & 106.74 & 30.66 & 19.35 & 0.114 & 1.223 & 0.883 & 1.336 \\
\cline{1-8}  
\multicolumn{8}{c}{\begin{minipage}[t]{1.\textwidth}
\caption{\label{tab:models} The values of basic parameters predicted by
different models}
\end{minipage}}
\\[-0.5cm]
\end{tabular}
\end{table} 
\end{center}
\ccc{
\begin{figure}
\begin{center}
\epsfig{file=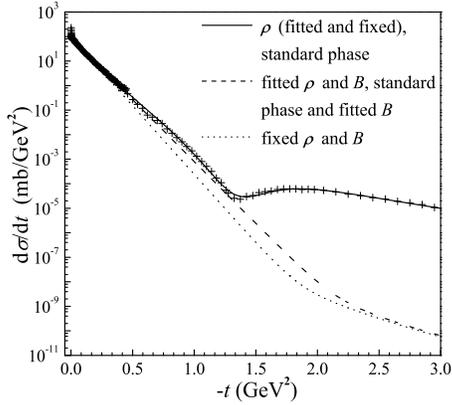,height=250pt,angle=0,width=250pt}
\caption{${{d \sigma} \over {dt}}$ for pp scattering at 53 GeV; all
graphs correspond to constant $\rho$.} 
\end{center} 
\end{figure}
}
\begin{figure}[h!]
\centerline{
\begin{tabular}{cc}
\includegraphics[width=0.4\textwidth]{prochfig.eps} & 
\includegraphics[width=0.4\textwidth]{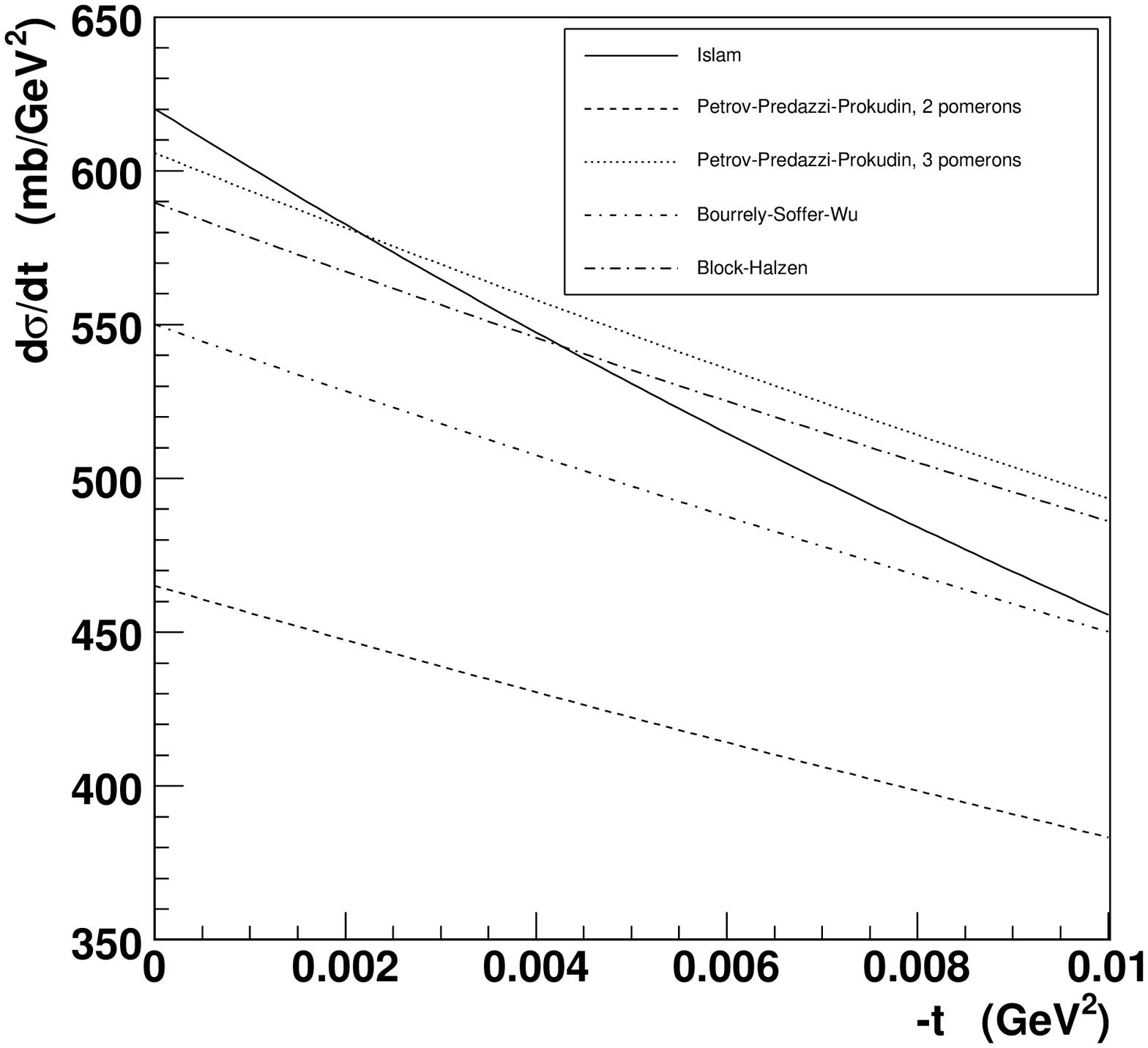}\\[-0.7cm]
\begin{minipage}[t]{0.48\textwidth}
\caption{\label{fig:prochfig}${{d \sigma} \over {dt}}$ for pp scattering at 53 GeV; all
graphs correspond to constant $\rho$.} 
\end{minipage} & \begin{minipage}[t]{0.45\textwidth}
\caption{\label{fig:elastic_dsigma_dt_0_1E-2}${{d \sigma} \over {dt}}$ predictions at low $|t|$
for pp scattering at 14 TeV corresponding to different models} 
\end{minipage}\\
&\\[-1.0cm]
\end{tabular}}
\end{figure}
\section{Model predictions for pp elastic scattering at the LHC}
\label{sec4}
In connection with the TOTEM experiment that will investigate 
the elastic pp scattering at energy of 14 TeV \cite{tdr}
we have studied the predictions of four models
proposed by the following authors: Islam, Luddy and Prokhudin 
\cite{isla}, Petrov, Predazzi and Prokhudin (with hadronic amplitude
corresponding to the exchange of two, resp. three pomerons -  
labelled as 2P, resp. 3P) \cite{petr}, Bourrely, Soffer 
and Wu  \cite{bour} and Block, Gregores, Halzen and Pancheri 
\cite{bloc}. All the given models have contained some free 
$s$ dependent parameters that have been established in these 
papers by fitting the experimental data on pp differential cross 
sections at several lower energies. Using these fitted values we have 
established the dynamical quantities: the total cross section,
momentum transfer distribution $ {{d \sigma} \over {dt}}$, 
the $t$ dependent diffractive slope $B(t)$ and the 
$t$ dependent quantity $\rho(t)$ at higher energy values. 
The values of $\sigma_{tot}$, $\sigma_{el}$, $B(0)$ 
and $\rho(0)$ for 14 TeV can be found in Table 1. 
The corresponding model predictions are shown
in Figs. 2 - 5. It is evident that the predictions 
of all the models differ rather significantly.
Fig. 2 shows different predictions for values of the 
differential cross sections at small $|t|$ values.
Great differences concern the values for total 
cross sections that are in direct relation 
to the values of differential cross 
section at $t = 0$; they run from 95 mb to 110 mb and differ rather
significantly from the value predicted by COMPETE collaboration \cite{cude}
$\sigma_{tot} \; = \; 111.5\;\; \pm \;\; 1.2 ^ {\;\;+4.1} _{\;\;-2.1}\;$ mb
which has been determined by extrapolation of the fitted lower 
energy data with the help of dispersion relations technique.
The predictions of ${d {\sigma} \over {dt}}$ values 
for higher values of $|t|$ are shown in Fig. 3.
Let us point out especially the second 
diffractive dip demonstrated by Bourrely, Soffer 
and Wu momentum transfer distribution.
The different predictions for $t$ dependence of the  
diffractive slopes $B(t)$ are shown in Fig. 4. 
Fig. 5 exhibits the $t$ dependence of the quantity $\rho (t)$.  
Table 1 contains also the values of the root-mean-squares calculated 
for each of the analyzed models with the formulas published in Ref. \cite{kun3}.
\ccc{
\begin{figure}
\begin{center}
\epsfig{file=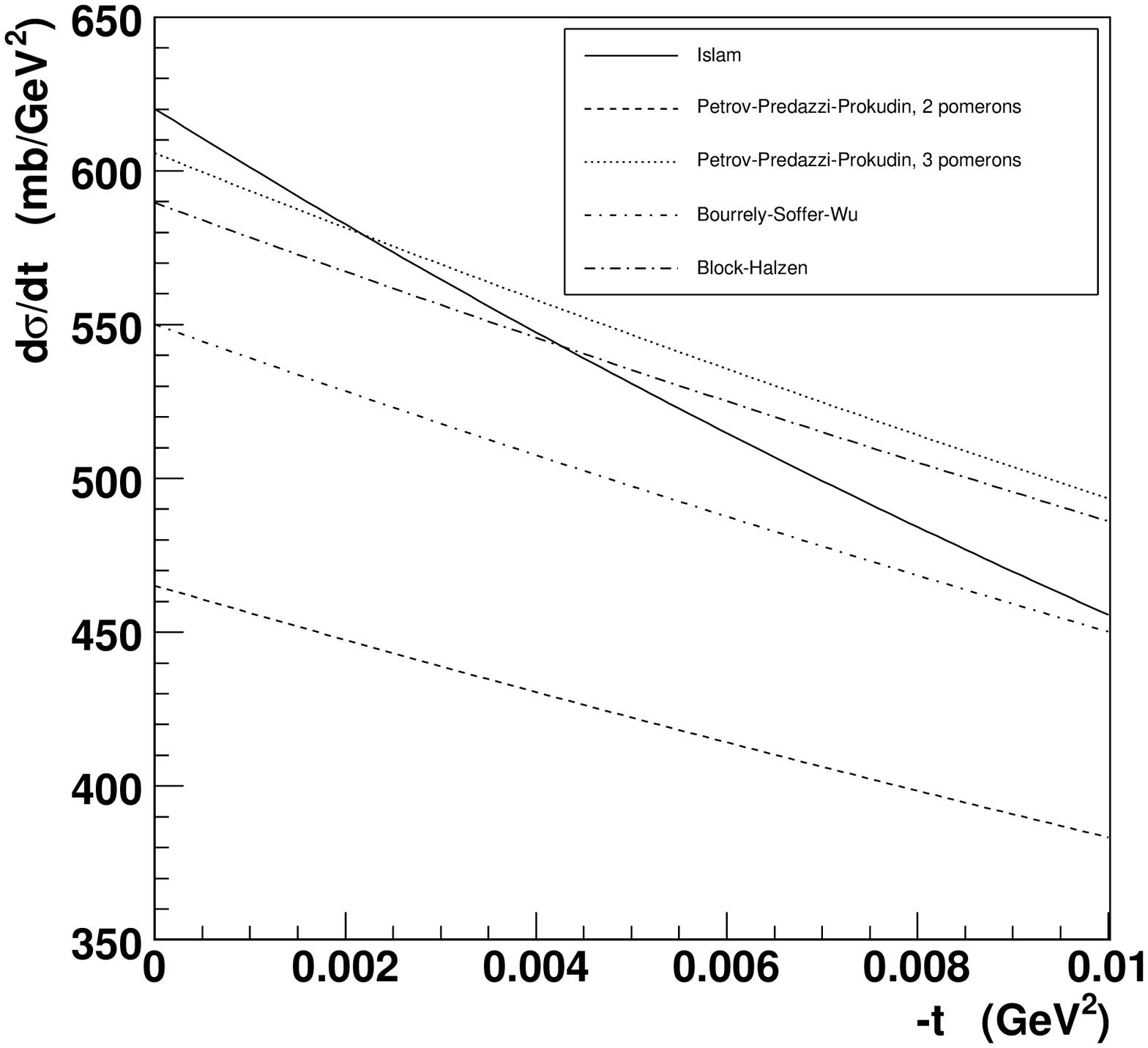,height=250pt,angle=0,width=250pt}
\caption{${{d \sigma} \over {dt}}$ predictions at low $|t|$
for pp scattering at 14 TeV corresponding to different models} 
\end{center} 
\end{figure}
}
\section{Luminosity estimation on the basis of pp elastic scattering 
at the LHC}
\label{sec5}
The accurate determination of the elastic amplitude is very
important in the case when the luminosity $\mathcal{L}$ is 
to be calibrated on the basis of elastic process; it holds 
that \cite{cahn} 
\begin{equation}
{1 \over \mathcal{L}} {d N_{el} \over {dt}} = { \pi \over {s p^2}}
 |F^{C+N} (s,t)|^2,
\label{lu1}
\end{equation}
where ${d N_{el} \over {dt}}$ is the counting rate
established experimentally at the given $t$. The so called
Coulomb calibration in the region of smallest $|t|$
where the Coulomb amplitude is dominant (reaching nearly
100 $\%$) can be hardly realized due to
technical limitations. The approach allowing to avoid
corresponding difficulties may be based on Eq.~(\ref{lu1}),
when the elastic counting rate can be, in principle, measured 
at any $t$ which can be reached, and the total elastic
scattering amplitude $F^{C+N}(s,t)$ may be determined 
with required accuracy at any $|t|$, too. Then the luminosity
$\mathcal{L}$ could be determined using Eq.~(\ref{lu1}). However,
\ccc{
\begin{figure}
\begin{center}
\epsfig{file=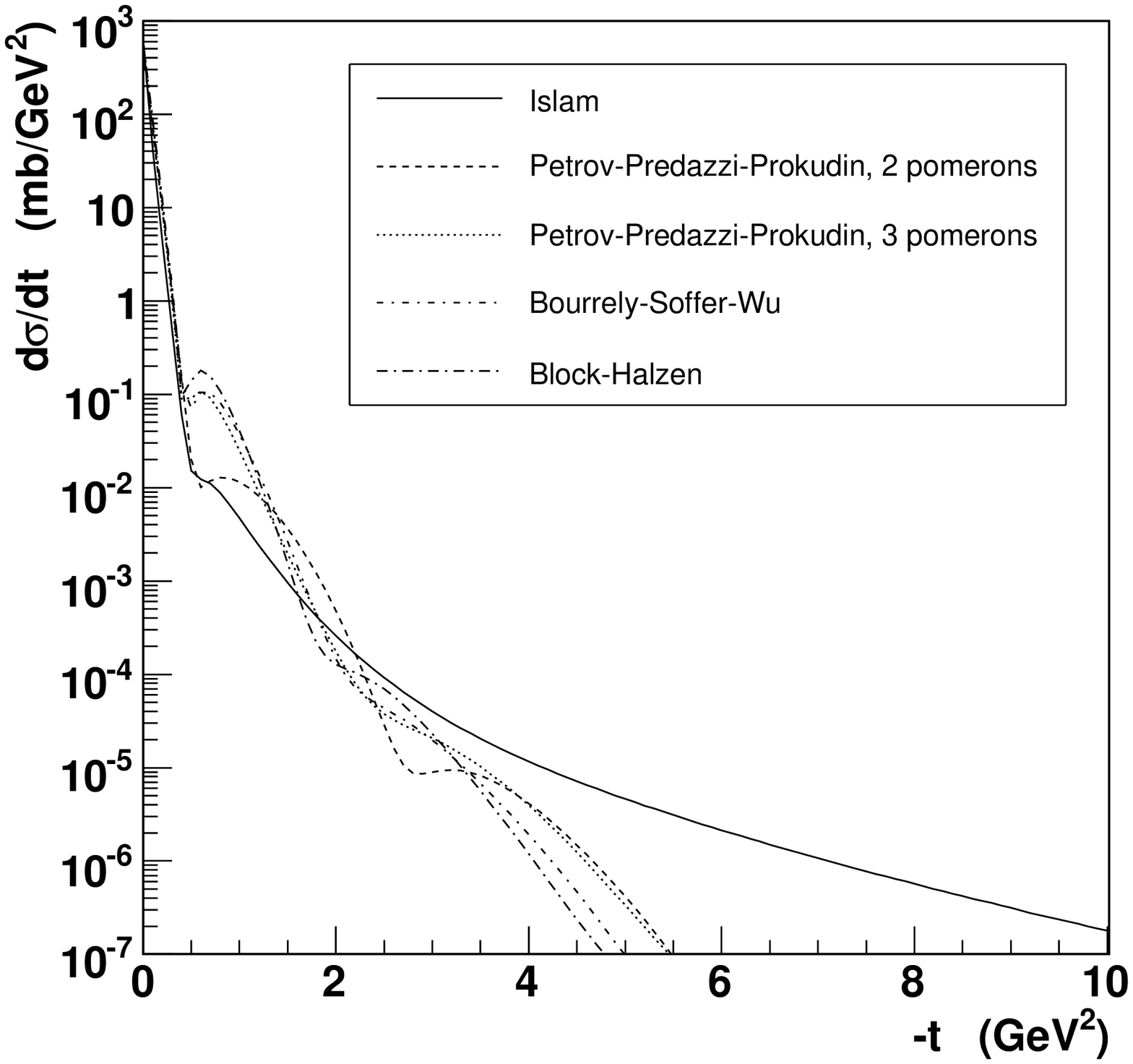,height=250pt,angle=0,width=250pt}
\caption{${{d \sigma} \over {dt}}$ predictions for pp scattering at 14 TeV
corresponding to different models} 
\end{center} 
\end{figure}
}
\ccc{
\begin{figure}
\begin{center}
\epsfig{file=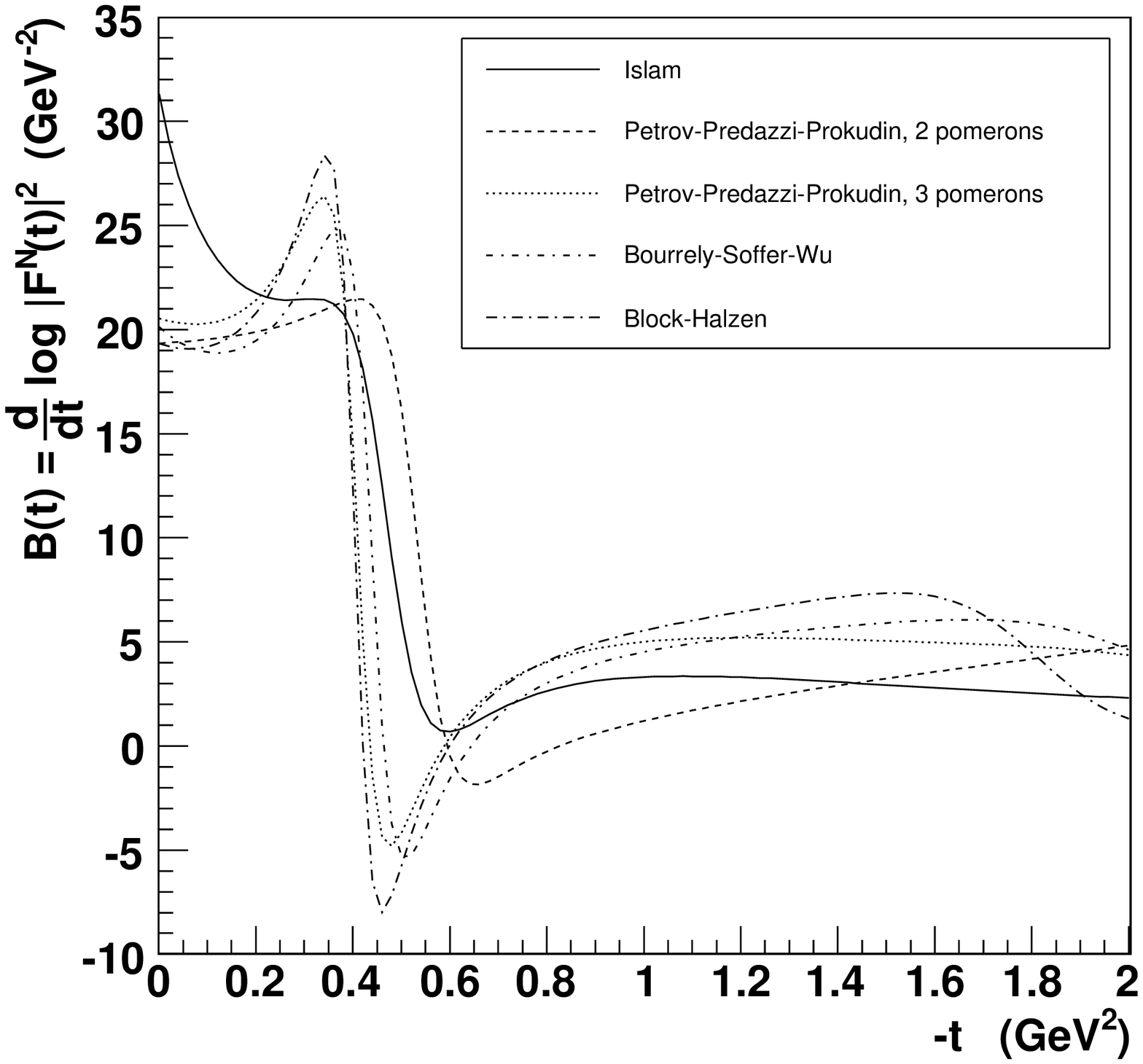,height=250pt,angle=0,width=250pt}
\caption{$t$ dependence of the slope predictions for pp scattering at 14 TeV
corresponding to different models} 
\end{center} 
\end{figure}
}
\ccc{
\begin{figure}
\begin{center}
\epsfig{file=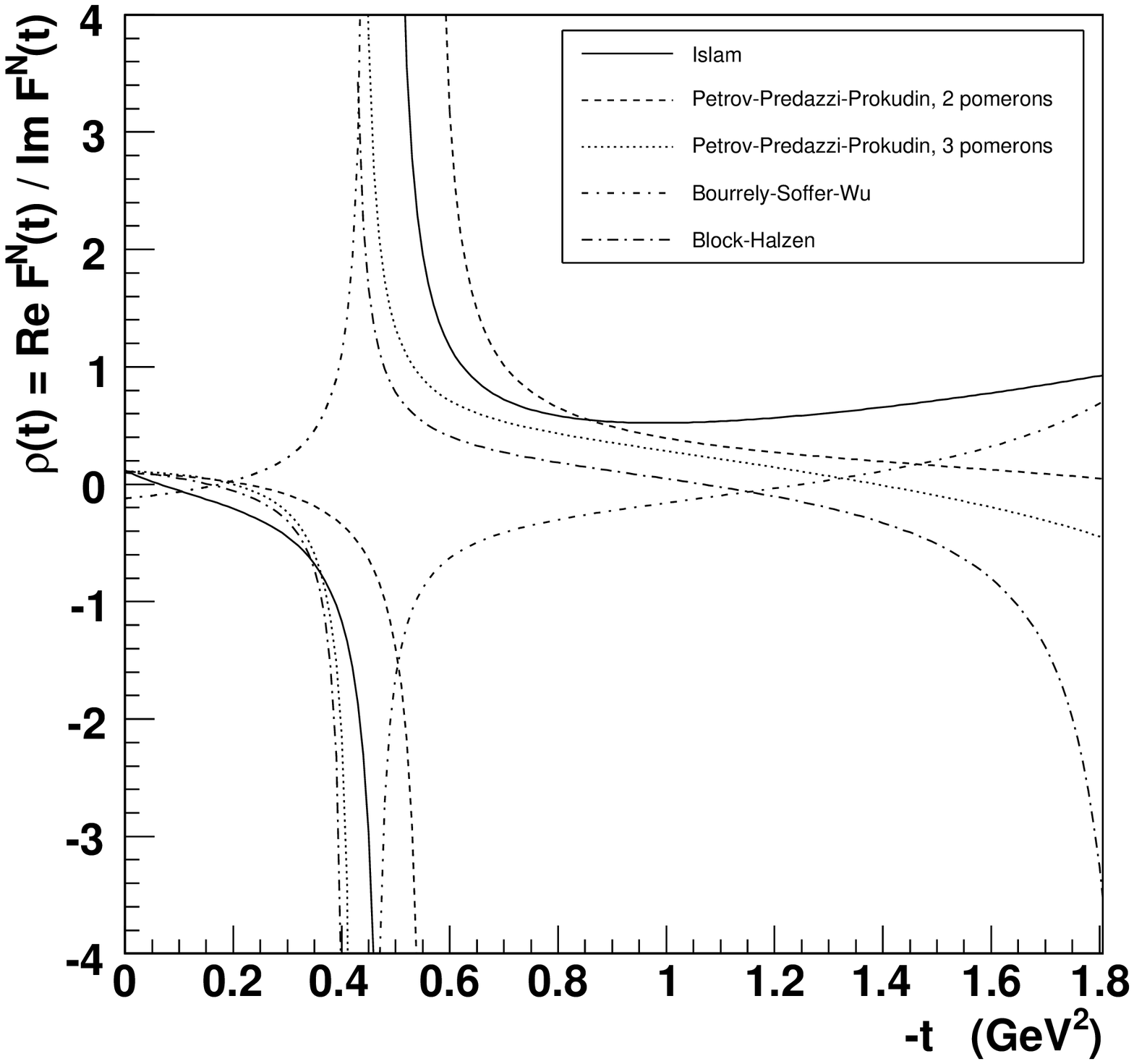,height=250pt,angle=0,width=250pt}
\caption{$\rho (t)$ predictions for pp scattering at 14 TeV
corresponding to different models} 
\end{center} 
\end{figure}
}
\begin{figure}[h!]
\centerline{
\begin{tabular}{cc}
\includegraphics[width=0.4\textwidth]{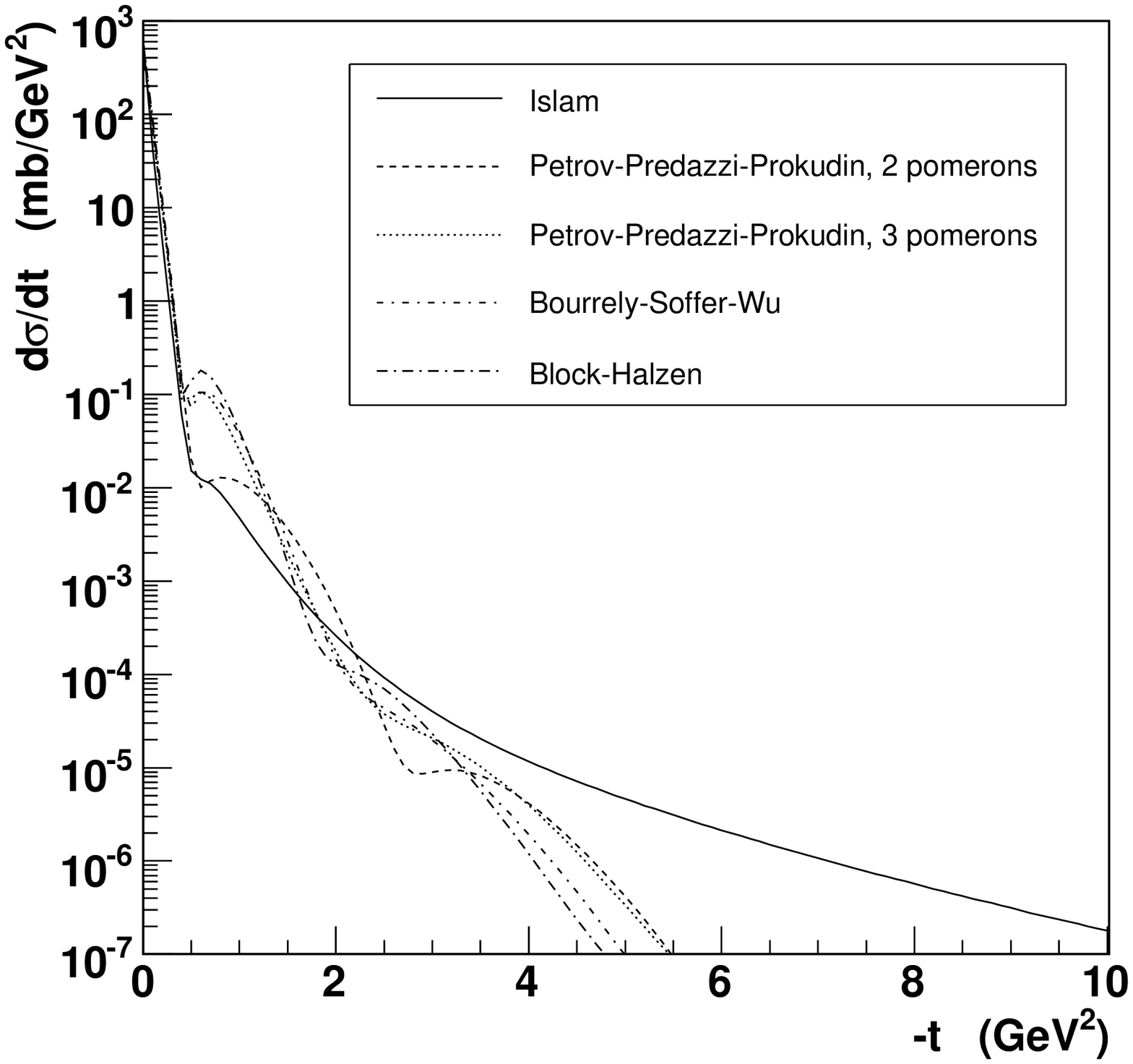} & 
\includegraphics[width=0.4\textwidth]{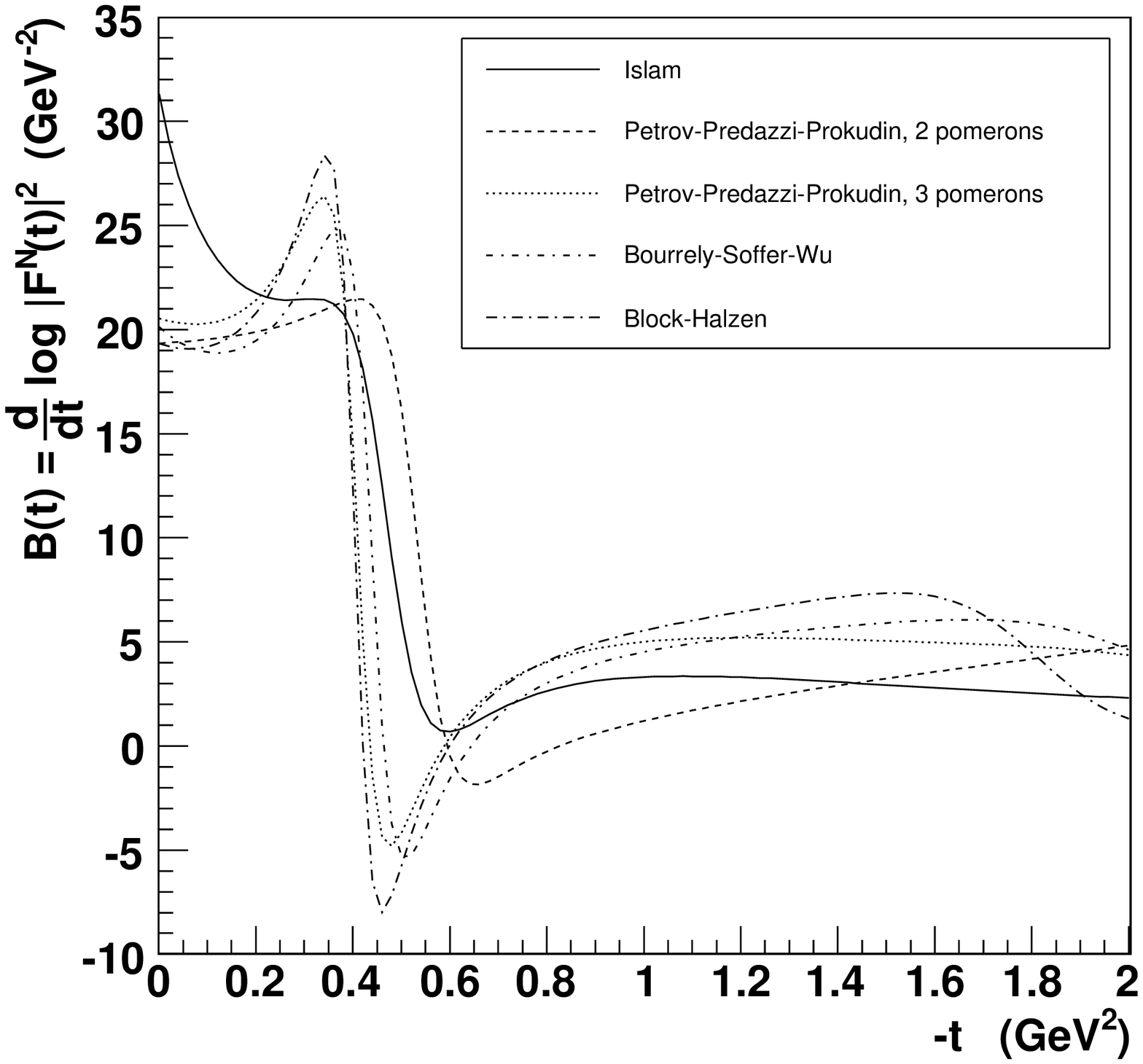}\\[-0.7cm]
\begin{minipage}[t]{0.45\textwidth}
\caption{\label{fig:elastic_dsigma_dt_0_10}${{d \sigma} \over {dt}}$ 
predictions for pp scattering at 14 TeV
corresponding to different models} 
\end{minipage} & \begin{minipage}[t]{0.45\textwidth}
\caption{\label{fig:elastic_slope_0_2_PH}$t$ dependence of the slope 
predictions for pp scattering at 14 TeV
for different models} 
\end{minipage}\\
&\\[-1.cm]
\end{tabular}}
\end{figure}
\begin{figure}[h!]
\centerline{
\begin{tabular}{cc}
\includegraphics[width=0.4\textwidth]{elastic_rho_0_2_PH.eps} & 
\includegraphics[width=0.4\textwidth]{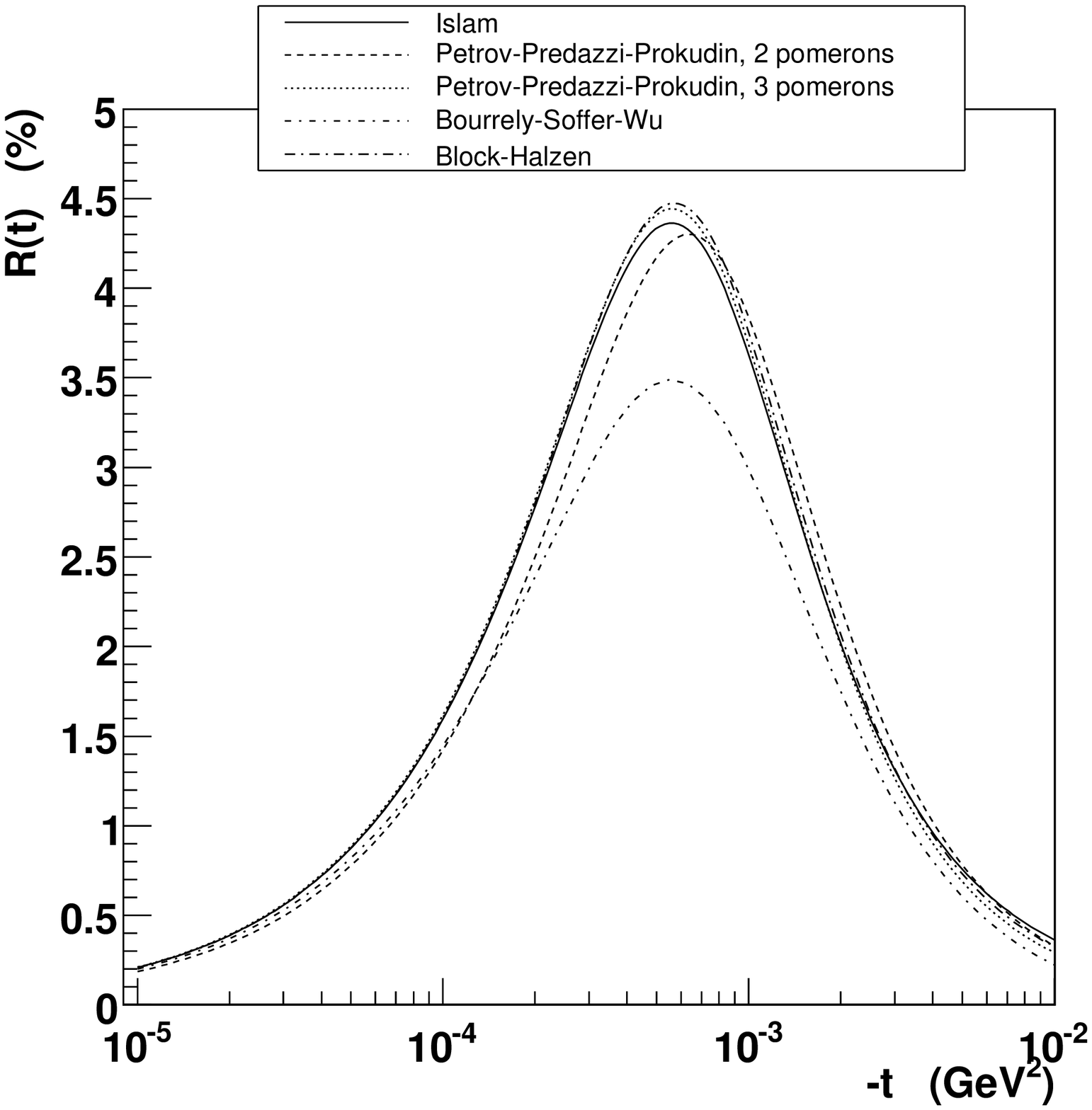}\\[-0.7cm]
\begin{minipage}[t]{0.45\textwidth}
\caption{\label{fig:elastic_rho_0_2_PH}$\rho (t)$ predictions for pp scattering at 14 TeV
corresponding to different models} 
\end{minipage} & \begin{minipage}[t]{0.45\textwidth}
\caption{\label{fig:elastic_R_1E-5_1E-2}$R(t)$ quantity predictions for pp scattering at 14 TeV
corresponding to different models} 
\end{minipage}\\
&\\[-1.cm]
\end{tabular}}
\end{figure}
in this case it is very important which formula 
for the total elastic amplitude $F^{C+N}(s,t)$ is made use of. 
In the following we will demonstrate possible differencies 
which can be obtained at different $t$ values in comparison with 
the standardly used West and Yennie approach. For this reason 
let us calculate the quantity
\begin{equation}
R(t) =  {{|F^{C+N}_{eik}(s,t)|^2 -
|F^{C+N}_{WY}(s,t)|^2}
\over {|F^{C+N}_{eik}(s,t)|^2}}, 
\label{lu2}
\end{equation}
where $F^{C+N}_{eik}(s,t)$ is the total elastic scattering 
eikonal model amplitude calculated for an investigated 
hadronic amplitude $F^N(s,t)$ while 
$F^{C+N}_{WY}(s,t)$ is the  West and Yennie 
total elastic scattering amplitude calculated for the 
same hadronic amplitude. The calculation was
performed for pp elastic scattering at the LHC 
energy of 14 TeV; in the former case Eqs.~(\ref{kl1}) 
- (\ref{kl3}) were used, in the latter one  
Eqs.~(\ref{wy5}) - (\ref{wy6}). The $t$ dependence of $R(t)$ 
for different models is shown in Fig. 6. The maximum 
values lie approximately at $t=-0.006$ GeV$^2$, showing that 
the differences of physically consistent eikonal models from 
West and Yennie formula may be almost 5 $\%$.
\ccc{
\begin{figure}
\begin{center}
\epsfig{file=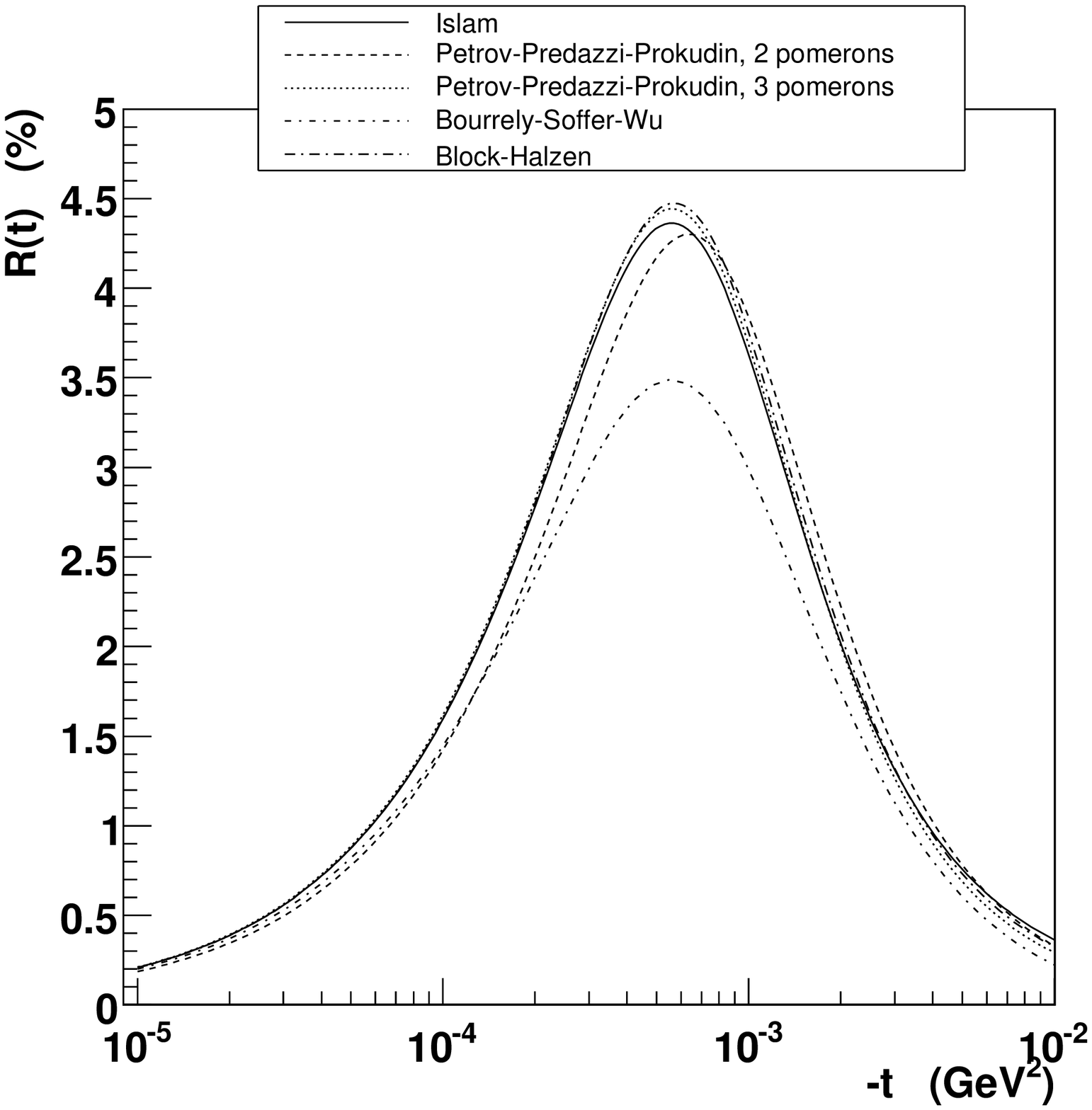,height=250pt,angle=0,width=250pt}
\caption{$R(t)$ quantity predictions for pp scattering at 14 TeV
corresponding to different models} 
\end{center} 
\end{figure}}
In the preceding case only the models with a weak
dependence of $\rho(t)$ on $t$ have been considered.
Yet larger difference may be obtained when the cases
with weak and strong dependences will be compared; i.e.,
the cases for central and peripheral distribution of elastic
hadron scattering - see the analysis of pp scattering at
ISR energies \cite{kun3}. It means that the luminosity
determined for the central and peripheral distributions of elastic 
pp scattering at LHC energy of 14 TeV may be burdened 
by a non-negligible mutual systematic error.
\vspace*{-0.25cm}


\begin{thebibliography}{99}
\bibitem{kun1}
V. Kundr\'{a}t and M. Lokaj\'{\i}\v{c}ek,
Phys. Lett. B 611 (2005) 102
\vspace*{-0.25cm}
\bibitem{vrko}
V. Kundr\'{a}t, M. Lokaj\'{\i}\v{c}ek and Ivo Vrko\v{c},
hep-ph/07060827
\vspace*{-0.25cm}
\bibitem{west}
G. B. West and D. Yennie, 
Phys. Rev. 172 (1968) 1413
\vspace*{-0.25cm}
\bibitem{amal}
U. Amaldi et al.,
Phys. Lett. 43B (1973) 231
\vspace*{-0.25cm}
\bibitem{kun2}
V. Kundr\'{a}t and M. Lokaj\'{\i}\v{c}ek,
Z. Phys. C 63 (1994) 619
\vspace*{-0.25cm}
\bibitem{proc}
J. Proch\'{a}zka, 
Bachelor Thesis, Charles University, Prague, May 2007
\vspace*{-0.25cm}
\bibitem{amos}
N. Amos et al.,
Nucl. Phys. B 262 (1985) 689
\vspace*{-0.25cm}
\bibitem{tdr}
V. Berardi et al, TOTEM Technical Design Report, 
CERN-LHCC-2004-002
\vspace*{-0.25cm}
\bibitem{isla}
M. M. Islam, R. J. Luddy and A. V. Prokhudin,
Phys. Lett. B 605 (2005) 115
\vspace*{-0.25cm}
\bibitem{petr}
V. A. Petrov, E. Predazzi and A. V. Prokhudin,
Eur. Phys. J. C28 (2003) 525
\vspace*{-0.25cm}
\bibitem{bour}
C. Bourrely, J. Soffer and T. T. Wu,
Eur. Phys. J. C28 (2003) 97
\vspace*{-0.25cm}
\bibitem{bloc}
M. M. Block, E. M. Gregores, F. Halzen and G. Pancheri,
Phys. Rev. D60   (1999) 0504024
\vspace*{-0.25cm}
\bibitem{cude}
J. R. Cudell et al., 
Phys. Rev. Lett. 89 (2002) 201801
\vspace*{-0.25cm}
\bibitem{kun3}
V. Kundr\'{a}t and M. Lokaj\'{\i}\v{c}ek,
Phys. Lett. B 544 (2003) 132
\vspace*{-0.25cm}
\bibitem{cahn}
M. M. Block and R. N. Cahn,
Rev. Mod. Phys. 57 (1985) 221
\end{thebibliography}
\end{document}